\documentclass[]{article}
\pdfoutput=1
\usepackage{arxiv}
\usepackage[hidelinks]{hyperref}
\usepackage{etoolbox}

\usepackage[american]{babel}
\usepackage[utf8]{inputenc}
\usepackage{graphicx}
\usepackage{booktabs}
\usepackage{xcolor}
\usepackage{listings}
\usepackage{multirow}
\usepackage{url}
\usepackage[nounderscore]{syntax}
\usepackage{microtype}
\usepackage{adjustbox}
\makeatletter
\def\MT@is@opt@char#1\iffontchar#2\char#3\else#4\fi\relax{%
  \MT@ifempty{#1}{%
    \iffontchar#2%
      \expandafter\chardef
        \csname\MT@encoding\MT@detokenize@c\@tempa\endcsname=#3\relax
    \fi
  }\relax
}
\makeatother

\usepackage{rotating}

\def\uschema{{\leavevmode\hbox{U-Schema}}}

\urldef{\mailsa}\path|{alberto.hernandez1,dsevilla,jmolina}@um.es|

\colorlet{punct}{red!60!black}
\definecolor{background}{HTML}{EEEEEE}
\definecolor{delim}{RGB}{20,105,176}
\colorlet{numb}{magenta!60!black}
\definecolor{comment}{RGB}{63,127,95} 
\definecolor{lightBlue}{RGB}{50,50,255}

\AtBeginEnvironment{grammar}{\scriptsize}
\lstdefinelanguage{ebnf}
{
  comment=[l]{//},
  morecomment=[s]{/*}{*/},
  morestring=[b]',
  morestring=[b]"
}

\lstdefinelanguage{athena}
{
  morekeywords=
  {
    FSet, Schema, Root, Entity, entity, Common, Variation, Identifier,
    String, List, Number, Ref, Aggr, Timestamp, Integer, as, Boolean, SQL,
    CREATE, TABLE, VARCHAR, NOT, NULL, FOREIGN, PRIMARY, KEY, REFERENCES,
    U, in, Import, Option
  },
  comment=[l]{//}
}

\lstdefinelanguage{alloy}
{
  morekeywords={pred, check, run, for, but, not, in, all, and, some, sig,set, or, True, False},
  comment=[l]{//},
  morecomment=[s]{/*}{*/},
  morestring=[b]',
  morestring=[b]"
}

\lstdefinelanguage{orion}
{
  morekeywords=
  {
    operations, Using, CAST, ATTR, TO, DELETE, ADAPT, ENTITY, PROMOTE, NEST,
    MORPH, AGGR, RENAME, ADD, REF, COPY, WHERE, CARD, rmId, rmEntity,
    Identifier, String, Number, AS, MULT, UNNEST, Double, Boolean, DELVAR,
    Timestamp
  },
  comment=[l]{//}
}

\lstdefinelanguage{mongodb}
{
  morekeywords=
  {
    \$jsonSchema, bsonType, required, properties, minimum, maximum, \$set,
    upsert, \$exists, \$unset, \$addFields, items, validator, description, enum,
    \$convert, \$rename, filter, updateMany, update, function, remove
  },
  comment=[l]{//},
  morecomment=[s]{/*}{*/},
  morestring=[b]',
  morestring=[b]"
}

\title{A Taxonomy of Schema Changes for NoSQL Databases\thanks{This
  work has been funded by the Spanish Ministry of Science, Innovation and
  Universities (project grant PID2020-117391GB-I00).}~$^,$\thanks{Formatted for
  arXiv.org.}}

\author{\href{https://orcid.org/0000-0002-1154-9192}{\includegraphics[scale=0.06]{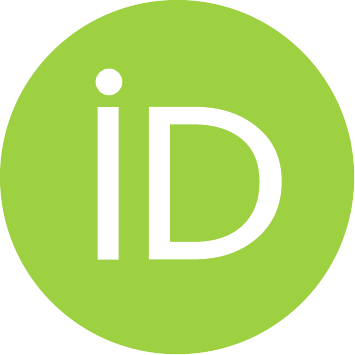}\hspace{1mm}Alberto Hernández Chillón}\\
  Faculty of Computer Science\\
  University of Murcia\\
  Murcia, Spain\\
  \texttt{alberto.hernandez1@um.es}\\
  \And \href{https://orcid.org/0000-0003-0551-8389}{\includegraphics[scale=0.06]{img/orcid.pdf}\hspace{1mm}Meike Klettke}\\
  Faculty of Computer Science and Data Science\\
  University of Regensburg\\
 Regensburg, Germany\\
  \texttt{meike.klettke@ur.de}\\
  \And \href{https://orcid.org/0000-0001-9313-008X}{\includegraphics[scale=0.06]{img/orcid.pdf}\hspace{1mm}Diego Sevilla Ruiz}\\
  Faculty of Computer Science\\
  University of Murcia\\
  Murcia, Spain\\
  \texttt{dsevilla@um.es}\\
  \And \href{https://orcid.org/0000-0003-4685-6659}{\includegraphics[scale=0.06]{img/orcid.pdf}\hspace{1mm}Jesús García Molina}\\
  Faculty of Computer Science\\
  University of Murcia\\
  Murcia, Spain\\
  \texttt{jmolina@um.es}\\
}
\renewcommand{\shorttitle}{A Taxonomy of Schema Changes for NoSQL Databases}

\begin{document}

\maketitle

\begin{abstract}
  Schema evolution is a crucial aspect in database management. The proposed
  taxonomies of schema changes have neglected the set of operations that
  involves relationships between entity types: aggregation and references,
  as well as the possible existence of structural variations for schema
  types, as most of NoSQL systems are schemaless. The distinction between
  entity types and relationship types, which is typical of graph schemas,
  is also not taken into account in the published works. Moreover, NoSQL
  schema evolution poses the challenge of having different data models, and
  no standard specification exists for them. In this paper, a generic
  approach for evolving NoSQL and relational schemas is presented, which is
  based on the \uschema{} unified data model that includes aggregation and
  reference relationships, and structural variations. For this data model,
  we introduce a taxonomy of schema changes for all the \uschema{}
  elements, which is implemented by creating the Orion database-independent
  language. We will show how Orion can be used to automatically generate
  evolution scripts for a set of NoSQL databases, and the feasibility of
  each schema operation will be analyzed through the performance results
  obtained. The taxonomy has been formally validated by means of Alloy, and
  two case studies show the application of Orion.
\end{abstract}


\twocolumn

\section{Introduction}\label{sec:introduction}

\begin{figure}[!ht]
  \centering
  \includegraphics[width=.5\textwidth]{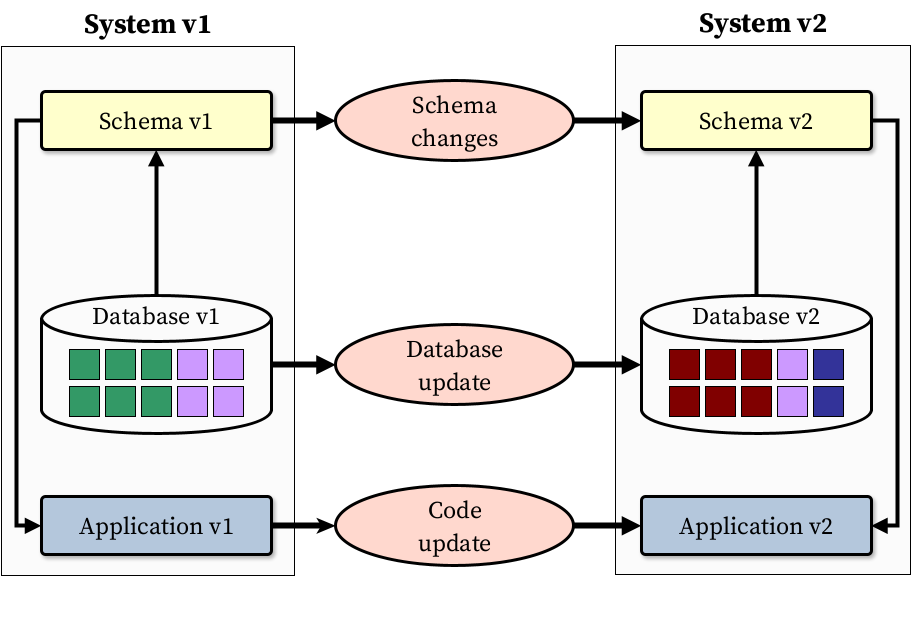}
  \caption{Data and code must be adapted when the schema changes.\label{fig:intro}}
\end{figure}

Database schemas have to be normally modified along the lifetime of
databases. These schema changes may be caused due to new functional or
non-functional requirements, or database refactoring, among other
situations. When this happens, stored data and application code must be
adapted to the new schema, as illustrated in Figure~\ref{fig:intro}. Thus,
the automation of schema changes is crucial to save effort and to avoid
data and application errors. For relational databases, a great deal of
research effort has been devoted to address the schema evolution problem,
and a number of tools, from prototypes to workbenches, have been built to
facilitate the schema change
management~\cite{curino-evolution2013,hick2003}. Also, schema evolution of
object databases has been extensively studied~\cite{xueli-databases1999}.
In this paper, we will address concerns related to schema evolution for
NoSQL (Not only SQL) stores.

Modern technologies as Web~2.0, smartphones and Internet of Things
motivated the building of applications with requirements that relational
systems could not meet. Then, NoSQL systems emerged to provide qualities as
flexibility to frequently change the schema, horizontal scalability and
availability, among others~\cite{fowler-nosql2012}. NoSQL systems are
commonly classified in four kinds of data models: columnar, document,
key-value and graph, but the lack of a standard or specification of the
data model leads to systems of the same category having different features.
Today, relational systems are still predominant but the idea of ``one size
does not fit all'' is widely accepted and polyglot
persistence~\cite{stonebraker-blog-2015} is considered the future trend, as
the growth of multi-model systems suggests: the first eight databases in
the DB-engines ranking\footnote{\url{https://db-engines.com/en/ranking}.}
are multi-model, and the main relational database vendors are including
support of NoSQL systems in their products.

To provide flexibility to adapt to schema changes, most of NoSQL systems
are ``schema-on-read,'' that is, the schema declaration is not required
prior to storing data (i.e.,~they are schemaless). However, this does not
mean that there is not a schema, but that it is implicit in data and code.
Stored data always are structured conforming to a schema, and code that
manipulates data is written according to that schema.

To achieve results similar to those obtained in relational systems, schema
evolution requires a larger research effort to overcome some limitations in
published approaches~\cite{klettke-bigdata2016,holubova-evolution2021}. For
example, schema change operations for relationships have not adequately
been addressed, and the existence of structural variations or relationship
types are not considered. Structural variation is possible for schemaless
systems since schemas guarantee that data conforms to a particular
structure, and relationship types are part of graph schemas. Moreover, the
building of automated solutions should be tackled for each NoSQL system.
The lack of a standard or specification for NoSQL data models makes this
task difficult.

In this work, we present a generic approach aimed to offer automated
support for NoSQL and relational systems. It is based on the \uschema{}
unified data model presented in~\cite{metamodel2021}. \uschema{} integrates
data models for the four kinds of NoSQL systems and the relational model,
and has more expressive power than other proposed generic data models
in~\cite{hainaut1994,cleve-typhon2020}, as discussed in detail
in~\cite{metamodel2021}. For example, \uschema{} distinguishes between
entity type and relationship type, includes aggregation and reference
relationships, and allows to represent the structural variation of the
schema types.

Our proposal consists of a taxonomy of \emph{schema change operations}
(SCOs) defined on \uschema{} and a domain-specific language (DSL) that
implements them. With this language, named \emph{Orion}, database
administrators and developers can write scripts that specify a set of SCOs.
From these scripts, the inferred or declared schema and the database are
automatically updated. The Orion engine has been built with data updaters
for several database systems, in particular MongoDB (document data model),
Cassandra (columnar), and Neo4j (graph). We have evaluated these data
updaters for two non-trivial case studies with real datasets: a refactoring
case to improve query performance, and an outlier migration problem. The
taxonomy has been formally validated by using the Alloy
language~\cite{jackson-alloy}. On the other hand, code updating is not
addressed here.

Our work contributes to the state of the art as follow:

\begin{itemize}
\item We have defined a taxonomy for NoSQL logical schemas, which includes
  a set of operations richer than those previously proposed. Being based on
  \uschema{}, we have considered changes on relationships, which are
  involved in frequent operations such as converting a particular reference
  into an aggregate or vice versa~\cite{coupal-patterns2019}, or copying
  features between schema types. Also, he have included changes related to
  variations, which could be very useful, e.g.,~joining all the variations
  in a single variation to remove outliers~\cite{klettke-schema2015}.
\item As far as we know, NoSQL data model heterogeneity has only been
  addressed in~\cite{cleve-typhon2020,holubova-evolution2019}. But
  taxonomies of these proposals do not include the changes above mentioned.
\item \emph{Orion} is a novel language to apply the operations of the
  taxonomy proposed. Orion data updaters have been built for three popular
  systems, and a study of the data updating cost for each operation has
  been performed.
\item Non-trivial case studies of schema evolution have been carried out by
  using real datasets.
\end{itemize}

Orion was already presented in the ``ER 2021--Conference on Conceptual
Modeling''~\cite{alberto-er2021}. Here, we show how that initial work has
been completed. An Orion data updater has been built for a graph store, in
particular Neo4j, and performance results have been obtained for each
schema change operation in our taxonomy. We have also used the Alloy formal
language to check the consistency of the operations, and have performed a
new case study. Below, the changes we have applied on the previous paper
are indicated:

\begin{itemize}
\item The performance analysis has been completely rewritten as a new
  experiment has been designed, and results for graph stores were not
  included in the ER paper.
\item A section has been added to describe the Alloy validation of the
  taxonomy.
\item Related work has been extended to consider the most promising
  research approaches regarding evolution taxonomies. Several criteria to
  compare approaches are proposed, and a table summarizes the comparison we
  have carried out.
\item A new case study is introduced to show the usefulness of our
  proposal: how outliers can be removed by changing the schema. The
  refactoring case study has been redesigned to be applied on a graph
  database instead of in a document database.
\end{itemize}

This paper has been organized in the following sections: The next Section
is used to introduce our data model. In Section~\ref{sec:taxonomy} we
define our abstract taxonomy of changes. In Section~\ref{sec:orion} the
Orion Language is described as a concrete implementation of that taxonomy.
Next, in Section~\ref{sec:casestudy} two case studies are discussed.
Section~\ref{sec:evaluation} shows a formal validation for the taxonomy and
a measure of performance of Orion. Section~\ref{sec:relatedwork} is used to
discuss related work, tools and research, and to compare the most promising
approaches. Finally, conclusions and future work are drawn in
Section~\ref{sec:conclusions}.


\section{\uschema{}: A Unified Data Model\label{sec:datamodel}}

\uschema{} is a generic metamodel that integrates the relational model and
data models for each of the four most common NoSQL paradigms: columnar,
document, key-value, and graph. A detailed description of \uschema{} is
presented in~\cite{metamodel2021}, where some of its applications are also
outlined. The use of different data models for different needs of
persistence is a trend, and \uschema{} was devised to build generic
database solutions. Here, \uschema{} is used to define a generic schema
evolution approach.

In this section, we will introduce \uschema{} through the \emph{Athena}
language~\cite{alberto-comonos2021}, which has been built to provide a
generic schema definition language with high expressive power. Although
most NoSQL systems are schemaless, this language is useful, for example,
when designing schemas from scratch, generating data for testing purposes,
or schema manipulation when there is no database whose schema can be
inferred.

\begin{figure}[!htb]
\begin{lstlisting}[language=athena]]
Schema Sales_department:1

Root entity Salesperson {
  Common {
    +id:          String,
    teamCode:     String,
    email:        String /^.+@.+\\.com$/,
    personalData: Aggr<PersonalData>&
  }
  Variation 1 {}
  Variation 2 {
    sales:        Aggr<SaleSummary>+,
    profits:      Integer (0 .. 9999)
  }
}

Entity PersonalData {
  city:           String,
  name:           String /^[A-Z][a-z]*$/,
  number:         Integer,
  street:         String,
  ? postcode:     Integer
}

Entity SaleSummary {
  saleId:         Ref<Sale>&,
  scheduledAt:    Timestamp,
  ? completedAt:  Timestamp,
  ? profits:      Integer
}

Root entity Sale {
  +id:            String,
  types:          List<String>,
  isActive:       Boolean,
  description:    String,
  profits:        Integer (0 .. 9999),
  exercises:      Ref<SeasonExercise as String>+
} + timeData

Root entity SeasonExercise {
  +id:            String,
  name:           String,
  description:    String,
  date_from:      Timestamp,
  date_to:        Timestamp
} + timeData

FSet timeData {
  createdAt:      Timestamp,
  updatedAt:      Timestamp
}
\end{lstlisting}%
\caption{The {\tt Sales department} schema defined using Athena.\label{fig:athena}}
\end{figure}

NoSQL data models can be classified in two
categories~\cite{fowler-nosql2012}: \emph{aggregate-based systems} where
aggregations prevail over references to connect objects, and \emph{graph
  systems} where only references are present. In graph systems, references
are instances of \emph{relationship types}, and aggregate-based systems
only have \emph{entity types}, and references are managed as
identifier-based joins. \uschema{} allows both kinds of NoSQL systems to be
represented.

A schema is formed by a set of \emph{schemas types} that can be
\emph{entity types} to represent domain entities or \emph{relationship
  types} to represent relationships between nodes in graph stores. In the
\texttt{Sales department} example (Figure~\ref{fig:athena}) there are five
entity types: \texttt{Salesperson}, \texttt{Sale}, \texttt{SeasonExercise},
\texttt{PersonalData}, and \texttt{SaleSummary}. The three former are root
entity types, that is, their objects are not embedded in any other object,
and the two latter are non-root as their instances are embedded into
\texttt{Salesperson} objects.

Because most of NoSQL systems are \emph{schema-on-read}, data objects of a
schema type can have different structure. Therefore, a schema type has a
set of structural variations in \uschema{}, which include a set of features
or properties. In Athena, when a schema type has variations, the features
are declared by separating those that are common to all variations from
those added by each particular variation. In the running example,
\texttt{Salesperson} is the only entity type with structural variations and
\texttt{Variation~1} only contains common features (\texttt{id},
\texttt{teamCode}, \texttt{email}, and \texttt{personalData}) while
\texttt{Variation~2} adds the \texttt{sales} and \texttt{profits} features.

A feature declaration specifies its name and type. There are four kinds of
features: \emph{keys}, \emph{attributes}, \emph{aggregates}, and
\emph{references}. \emph{Attributes} and \emph{aggregates} denote the
features that hold the values of a database object. For attributes, the
type can be either scalar (\emph{Number}, \emph{String}, \emph{Boolean},
etc) or structured (\emph{Set}, \emph{List}, \emph{Map}, and \emph{Tuple}).
In the case of aggregates, the type is a non-root entity type. For example,
\texttt{Salesperson.email} is an attribute of type String, and
\texttt{Salesperson.personalData} specifies that \texttt{PersonalData}
objects are embedded in \texttt{Salesperson} objects.

\emph{Keys} and \emph{references} are features that hold an object
identifier. They are formed by one or more attributes. Keys are declared
with the modifier ``\texttt{+}'', and references are associated to a root
type. For example, \texttt{Salesperson.id} is a key, and
\texttt{Sale.exercises} specifies that \texttt{Sale} objects reference
\texttt{SeasonExercise} objects whose key is the \texttt{id} attribute. A
cardinality needs to be specified for references and aggregations, such as
\emph{one to one} or \emph{one to many}. The features may also have
modifiers as \emph{key}~(``\texttt{+}'') or
\emph{optional}~(``\texttt{?}'').


\section{Taxonomy of Changes for NoSQL Databases\label{sec:taxonomy}}

In schema evolution approaches, the set of changes that can be applied on a
particular data model are usually organized in form of a
taxonomy~\cite{banerjee-ooevolution1987,holubova-evolution2021}. Several
categories are established depending on the kind of schema element affected
by a change. Here, we present a taxonomy for the \uschema{} data model
introduced in the previous section, which includes operations for all its
elements. In this way, our taxonomy includes all the operations proposed in
the studied taxonomies, and adds new operations, such as those related to
aggregates, references, relationship types, and variations, as shown later.

Next, the terminology used to define the semantics of operations in our
taxonomy is introduced. Let $T$ be the set of schema types, and let $E$ be
the set of entity types $E=\{E_i\}, i=1 \ldots n$, $T = E$ in the case of
aggregate-based stores, while $T = E \cup R$, where
$R=\{R_i\}, i=1 \ldots m$ denotes the set of relationship types, in the
case of graph stores. Each schema type $t \in T$ includes a set of
structural variations $V^t = \{v^t_1, v^t_2, \ldots ,v^t_n\}$, with
$v^{t}_{i}.features$ denoting the set of features of a variation
$v^{t}_{i}$. Then, the set of features of a schema type $t$ is
$F^t=\bigcup_{i=1}^n v^t_i.features$, which will include attributes,
aggregates, and references, and $C^t \subset F^t$ denotes the set of common
features of a type $t$. We will use \emph{dot notation} to refer to parts
of a schema element, e.g.,~given an entity type $e$, $e.name$ and
$e.features$ refer to the name and set of features ($F^e$), respectively,
of the entity type.

The proposed taxonomy is shown in Table~\ref{tab:taxonomy}. In a similar way
to~\cite{curino-evolution2013}, we have added operations taking into account a
compromise between atomicity, usability, and reversibility. In the case of
changes affecting variations, usefulness and atomicity have prevailed on
reversibility. Each SCO is defined by an identifying name, together with
information regarding the gaining or loss of information the operation causes on
the schema, denoted by a $C^s$ notation as follows:

\begin{itemize}
\item $C^+$ if an operation carries an \emph{additive change},
  e.g.,~\emph{Add Schema Type}.
\item $C^-$ if a \emph{subtractive change} occurs, e.g.,~\emph{Delete
    Feature}.
\item $C^{+,-}$ denotes an operation in which there is a gain and a loss of
  information, e.g.,~\emph{Move Feature}.
\item $C^=$ means no change in information, e.g.,~\emph{Rename Schema
    Type}.
\item $C^{+|-}$ adds or subtracts information, depending on the operation
  parameters, e.g.,~\emph{casting} a feature to boolean.
\end{itemize}

As noted in~\cite{curino-evolution2013}, a schema change operation can be
considered a function whose input is a schema $S$ and a database $D$
conforming to it and produces as output a modified schema $S^\prime$ and
the database $D^\prime$ that results of updating $D$ to conform to
$S^\prime$. In this paper, the schema operation semantics is defined in
form of pre and postconditions, which appear in the second and third
column. Note that the postconditions only specify the changes on the schema
and not the changes on the database, because these depend on the concrete
data model. Since the operations semantics would be expressed very
similarly to specifying the database change, this semantics is not included
here, but we have added a comment to the postcondition of the \emph{Adapt}
operation to show that its effect is different to the \emph{Delvar}
operation.

As Table~\ref{tab:taxonomy} shows, taxonomy operations are classified in~6
categories that correspond to \uschema{} elements: \emph{schema types},
\emph{variations}, \emph{features}, \emph{attributes}, \emph{references},
and \emph{aggregates}.

\begin{table*}[!ht]
{%
\caption{Schema Change Operations of the Taxonomy.\label{tab:taxonomy}}
\begin{adjustbox}{width=1\textwidth}
\tiny\sffamily%
\noindent%
\begin{tabular}{p{.15\textwidth}p{.45\textwidth}p{.35\textwidth}}
\toprule
  & \multicolumn{1}{c}{\bfseries Precondition} & \multicolumn{1}{c}{\bfseries Postcondition} \\
\midrule
  \multicolumn{3}{l}{\begin{tabular}[l]{@{}l@{}}{\bfseries Schema Type Operations}\\{\bfseries (Entity Type and Relationship Type)}\end{tabular}}\\
\midrule

{\bfseries Add} \hfill ($C^+$)
  & Let $t$ be a new schema type, $t \notin T$
  & $t \in T$ \\

{\bfseries Delete} \hfill ($C^-$)
  & Given a schema type $t \in T$
  & $t \notin T$ \\

{\bfseries Rename} \hfill ($C^=$)
  & Given a schema type $t \in T$ and a string value $n$, $n \notin T.names$
  & $t.name = n$ \\

{\bfseries Extract} \hfill ($C^{+,=}$)
  & Given a schema type $t \in T$, a set of features $fs \subset F^{t}$ and a string value $n \notin T.names$
  & $t \in T \land t_1 = T.new \land t_{1}.name = n \land t_{1}.features = fs$ \\

{\bfseries Split*} \hfill ($C^=$)
  & Given a schema type $t \in T$, two sets of features $fs_1 \subset F^t \land
  fs_2 \subset F^t$ and two string values $n_1,n_2 \notin T.names$
  & $t \notin T \land t_1 = T.new \land t_2 = T.new \land t_1.name = n_1 \land t_1.features = fs_1 \land t_2.name = n_2 \land t_2.features = fs_2$ \\

{\bfseries Merge} \hfill ($C^=$)
  & Given two schema types $t_1,t_2 \in T$ and a string value $n \notin T.names$
  & $t_1,t_2 \notin T \land t = T.new \land t.name = n \land t.features = t_1.features \cup t_2.features$ \\

\midrule
  \multicolumn{3}{l}{\bfseries Structural Variation Operations} \\
\midrule

{\bfseries Delvar} \hfill ($C^-$)
  & Given a schema type $t \in T$ and a variation $v^t \in V^{t}$
  & $v^t \notin V^t$ \\

{\bfseries Adapt} \hfill ($C^=$)
  & Given a schema type $t \in T$ and two variations $v_1^t, v_2^t \in V^{t}$
  & $v_1^t \notin V^t$\hfill\emph{(Data is migrated from $v_1^t$to $v_2^t$)}\\

{\bfseries Union} \hfill ($C^+$)
  & Given a schema type $t \in T \land V^t \neq \{\}$
  & $ V^t = \{v_m\} \land v_m.features = \cup_{i=1}^n v_i^t.features$ \\

\midrule
  \multicolumn{3}{l}{\begin{tabular}[l]{@{}l@{}}{\bfseries Feature Operations}\\{\bfseries (Attribute, Reference and Aggregate)}\end{tabular}}\\
\midrule

{\bfseries Delete} \hfill ($C^-$)
  & Given a schema type $t \in T$ and a feature $f \in F^t$
  & $f \notin F^t$ \\

{\bfseries Rename} \hfill ($C^=$)
  & Given a schema type $t \in T$, a feature $f \in F^t$, and a string value $n \notin t.features.names$
  & $f.name = n$ \\

{\bfseries Copy} \hfill ($C^+$)
  & Given two schema types $t_1,t_2 \in T$ and a feature $f \in F^{t_1} \land f \notin F^{t_2}$
  & $f \in F^{t_1} \land f \in F^{t_2}$ \\

{\bfseries Move*} \hfill ($C^{+,-}$)
  & Given two schema types $t_1,t_2 \in T$ and a feature $f \in F^{t_1} \land f \notin F^{t_2}$
  & $f \notin F^{t_1} \land f \in F^{t_2}$ \\

{\bfseries Nest} \hfill ($C^{+,-}$)
  & Given an entity type $e_1 \in E$, a feature $f \in F^{e_1}$, and an aggregate $ag \in F^{e_1} \land ag.type = e_2 \land f \notin F^{e_2}$
  & $f \notin F^{e_1} \land f \in F^{e_2}$ \\

{\bfseries Unnest} \hfill ($C^{-,+}$)
  & Given an entity type $e_1 \in E$, an aggregate $ag \in F^{e_1} \land ag.type = e_2$, and a feature $f \notin F^{e_1} \land f \in F^{e_2}$
  & $f \in F^{e_1} \land f \notin F^{e_2}$ \\

\midrule
  \multicolumn{3}{l}{\bfseries Attribute Operations} \\
\midrule

{\bfseries Add} \hfill ($C^+$)
  & Given a schema type $t \in T$, let $at$ be an attribute, $at \notin C^t$
  & $at \in C^t$ \\

{\bfseries Cast} \hfill ($C^{+|-}$)
  & Given a schema type $t \in T$, an attribute $at \in F^t$, and a scalar type $st$
  & $at.type = st$ \\

{\bfseries Promote} \hfill ($C^=$)
  & Given an entity type $e \in E$ and an attribute $at \in F^e \land at.key = False$
  & $at.key = True$ \\

{\bfseries Demote} \hfill ($C^=$)
  & Given an entity type $e \in E$ and an attribute $at \in F^e \land at.key = True$
  & $at.key = False$ \\

\midrule
  \multicolumn{3}{l}{\bfseries Reference Operations} \\
\midrule

{\bfseries Add} \hfill ($C^+$)
  & Given a schema type $t \in T$, let $rf$ be an reference, $rf \notin C^t$
  & $rf \in C^t$ \\

{\bfseries Cast} \hfill ($C^{+|-}$)
  & Given a schema type $t \in T$, a reference $rf \in F^t$, and a scalar type $st$
  & $rf.type = st$ \\

{\bfseries Mult} \hfill ($C^{+|-}$)
  & Given a schema type $t \in T$, a reference $rf \in F^t$, and a tuple $(l,u) \in \{(0, 1), (1, 1), (0, -1), (1, -1)\}$ & $rf.lowerBound = l \land rf.upperBound = u$ \\

{\bfseries Morph} \hfill ($C^=$)
  & Given a schema type $t \in T$ and a reference $rf \in F^t$, let $ag$ be a new aggregate, $ag \notin F^t$
  & $rf \notin F^t \land ag \in F^t \land ag.name = rf.name \land ag.type = rf.type$ \\

\midrule
  \multicolumn{3}{l}{\bfseries Aggregate Operations} \\
\midrule

{\bfseries Add} \hfill ($C^+$)
  & Given an entity type $e \in E$, let $ag$ be an aggregate, $ag \notin C^e$
  & $ag \in C^e$ \\

{\bfseries Mult} \hfill ($C^{+|-}$)
  & Given an entity type $e \in E$, an aggregate $ag \in F^e$ and a tuple $(l,u) \in \{(0, 1), (1, 1), (0, -1), (1, -1)\}$ & $ag.lowerBound = l \land ag.upperBound = u$ \\

{\bfseries Morph} \hfill ($C^=$)
  & Given an entity type $e \in E$ and an aggregate $ag \in F^e$, let $rf$ be a new reference, $rf \notin F^e$
  & $ag \notin F^e \land rf \in F^e \land rf.name = ag.name \land rf.type = ag.type$ \\

\bottomrule
\end{tabular}%
\end{adjustbox}
}
\end{table*}

The \emph{Schema type} category groups operations that can be applied
indistinctly on \emph{entity} and \emph{relationship types}. In addition to
the atomic operations: \emph{Add}, \emph{Delete}, and \emph{Rename}, three
complex operations are added to create new schema types. The \emph{Extract}
operation creates a new schema type by copying some of the features of an
existing schema type, and leaving the original schema type unmodified. The
\emph{Split} operation divides an existing schema type into two new schema
types by separating its features into two subsets, and the original schema
type ceases to exist. The \emph{Merge} operation can be understood as the
inverse of the previous operation: a new schema type is created as the
union of two existing ones, which are removed afterwards.

The \emph{Structural Variations} category groups three operations defined
to manipulate them. The \emph{Delvar} operation deletes a given variation,
\emph{Adapt} deletes a given variation but also migrates data belonging to
the deleted variation to a new variation, and \emph{Union} joins all the
variations of a schema type into a single one. The first two operations
could be useful, for example, to remove outliers (i.e.,~variations with a
small number of elements)~\cite{klettke-schema2015}. Since \emph{Delvar}
and \emph{Adapt} cause the same changes on the schema (a variation is
deleted), a comment has been added in the \emph{Adapt} postcondition to
indicate the effect on the database.

Similarly to the \emph{Schema Type} category, the \emph{Feature} category
groups the operations with the same semantics for attributes, aggregates,
and references. It includes operations to~(i)~copy a feature from a schema
type to another one, either maintaining (\emph{Copy}) or not (\emph{Move})
the feature copied in the original schema type; and~(ii)~move a feature
from/to an aggregate: \emph{Nest} and \emph{Unnest}.

The \emph{Attribute} category includes operations to \emph{Add} a new
attribute, change its type (\emph{Cast}), and add/remove an attribute
to/from a key: \emph{Promote} and \emph{Demote}. The \emph{Reference}
category includes the \emph{Add} and \emph{Cast} operations commented for
attributes, \emph{Mult} to change the multiplicity, and the \emph{Morph}
operation to transform a reference to an aggregate. Finally the
\emph{Aggregate} category includes operations \emph{Add}, \emph{Mult} and
\emph{Morph} commented for references. Please note that there is not
\emph{Key} category because in \uschema{} a \emph{Key} is a \emph{logical
  feature} that is always bound to an attribute, and therefore keys can be
created and deleted by means of the attribute operations \emph{Add},
\emph{Promote}, and \emph{Demote}.

All the listed SCOs, except for\emph{Split} and \emph{Move}, are
\emph{atomic operations}. This means that these basic SCOs cannot be
implemented as a combination of two or more other SCOs. On the other hand,
\emph{Split} and \emph{Move} are \emph{non-atomic operations} because they
can be implemented by using other SCOs (\emph{Movie} is composed of a
\emph{Copy} and \emph{Delete} feature operations, and \emph{Split} can be
defined as two \emph{Extract} and a \emph{Delete} schema type operations).
These two operations have still been added to the taxonomy because they are
recurrent operations in refactoring scenarios, and other approaches have
considered them.


\section{Implementing the Taxonomy in Orion\label{sec:orion}}

Orion is the language created to implement the taxonomy defined on
\uschema{}. With Orion, developers can declare and execute change
operations in a system-independent way. Metamodeling has been applied to
define the language: a metamodel expresses the abstract syntax, and
notation or concrete syntax and semantics are defined on the
metamodel~\cite{brambilla2012}. The Orion metamodel specifies the Orion
grammar as an Ecore model~\cite{steinberg-emf2009}, i.e.,~an
object-oriented domain model. Here, the metamodel is not shown because we
consider the grammar notation is enough to understand the contribution of
the language. The notation will be explained by showing examples for the
running example, and, finally, semantics will be illustrated by indicating
the generated code.

\subsection{Concrete Syntax}

SCOs can be easily expressed like commands of a command-line language.
Therefore, the syntax of Orion is very simple as illustrated in
Figure~\ref{fig:orion_ebnf} in which an excerpt of its EBNF grammar is
shown. Note that the general format for the majority of operations is a
keyword denoting the change operation (e.g.,~\emph{Add} or \emph{Delete})
followed by another keyword to indicate the kind of schema element it
affects (e.g.,~\emph{Entity} or \emph{Relationship}, \emph{Aggregate} or
\emph{Reference}), and finally a list of arguments. The Orion syntax has
been defined to let operations be written as concise as possible, e.g.,~it
is possible to apply certain operations over all schema types by using the
``\texttt{*}'' wildcard, as in \texttt{DELETE~*::name}, and operations can
define a list of parameters as in \texttt{DELETE Sales::types, isActive,
  description}. Operations can also be applied to specific variations of a
schema type, as in \texttt{RENAME~*(v1,v3)::phone TO newPhone}. Differences
between aggregate-based systems and graph systems are expressed through
optional parameters in some operation commands, e.g.,~\texttt{ADD~REF}
requires to indicate a target entity type and a \emph{join} condition, and
has two optional parameters: the primitive type of the references values in
the case of an aggregated-based store, and a set of attributes for a graph
store.

\begin{figure}[!ht]
\hrulefill
\begin{grammar}
<RenameOp>    ::= `RENAME' `ENTITY' <RenameSpec>

<RenameSpec>  ::= <EName> `TO' <EName>

<AdaptOp>     ::= `ADAPT' `ENTITY' <EName> `::' `v' <VarId> `TO' `v' <VarId>

<DeleteFeatOp> ::= `DELETE' <MultipleFSelector>

<RenameFeatOp> ::= `RENAME' <SingleFSelector> `TO' <QName>

<NestFeatOp> ::= `NEST' <MultipleFSelector> `TO' <QName>

<CastAttrOp> ::= `CAST' `ATTR' <MultipleFSelector> `TO' <PrimitiveType>

<PromoteAttrOp> ::= `PROMOTE' `ATTR' <MultipleFSelector>

<AddRefOp> ::= `ADD' `REF' <SingleFSelector> `:' (<PrimitiveType> | `{' ( <SimpleDataF> ( `,' <SimpleDataF> )* )? `}' ) (`?' | `\&' | `*' | `+' ) `TO' <EName> (`WHERE' <ConditionDecl>)?

<MultipleFSelector> ::= ( <EName> ( `(' <VarId> ( `,' <VarId> )* `)' )?
                        \alt `*' )
                        `::' <QName> ( `,' <QName> )*

<SingleFSelector> ::= ( <EName> ( `(' <VarId> ( `,' <VarId> )* `)' )?
                      \alt `*' )
                      `::' <QName>

<ConditionDecl>   ::= <QName> `=' <QName>

<QName>           ::= ID ( `.' ID )*

<EName>           ::= ID

<VarId>           ::= INT
\end{grammar}
\hrulefill
\caption{EBNF excerpt of the Orion language.\label{fig:orion_ebnf}}
\end{figure}

Figure~\ref{fig:orion1} shows an Orion script that applies changes on the
{\tt Sales department} schema of Figure~\ref{fig:athena}. An Orion script
starts with a \texttt{Using} statement that indicates the schema on which
the changes are applied. This declaration allows runtime checking of the
validity of each operation on the current schema. Note that the schema has
to be updated after the execution of each operation of the script, so that
the checking can be correctly performed. The operations are therefore
sequentially executed.

The script of Figure~\ref{fig:orion1} shows changes on several entity types
of the schema, illustrating most of the different changes in the taxonomy:
casting on attributes (\texttt{*::profits},
\texttt{PersonalData::postCode}, and \texttt{SaleSummary::isCompleted}),
deleting attributes (\texttt{Sale::isActive}), nesting attributes to an
aggregate (\texttt{Salesperson::email} and \texttt{PrivateData::city,
  postcode, street}), morphing an aggregate to a reference
(\texttt{Salesperson::personalData}), renaming entity types
(\texttt{Salesperson}) and features (\texttt{SaleSummary::completedAt}),
and adapting a variation (\texttt{Salesperson::v1}). Operations to create
entity types and aggregates are also shown (\texttt{Company},
\texttt{Company::media} and \texttt{PersonalData::address}).

\begin{figure}[!ht]
\begin{lstlisting}[language=orion]
Sales_ops operations
Using Sales_department:1

// Sale operations
CAST ATTR *::profits TO Double
DELETE Sale::isActive

// PersonalData operations
CAST ATTR PersonalData::postcode TO String
ADD AGGR PersonalData::address:{country:String}& AS Address
NEST PersonalData::city, postcode, street TO address

// Salesperson operations
ADAPT ENTITY Salesperson::v1 TO v2
NEST Salesperson::email TO personalData
MORPH AGGR Salesperson::personalData TO privateData
RENAME ENTITY Salesperson TO Employee

// SaleSummary operations
RENAME SaleSummary::completedAt TO isCompleted
CAST ATTR SaleSummary::isCompleted TO Boolean
RENAME ENTITY SaleSummary TO Summary

// Adding a new type
ADD ENTITY Company: { +id: String, code: String,
              name: String, numOfEmployees: Number }

// Adding new features
PROMOTE ATTR Company::code
ADD AGGR Company::media: { twitterProf: String,
  fbProf: String, webUrl: String, ytProf: String}& TO Media
\end{lstlisting}%
\caption{Refactoring of the {\tt Sales department} schema using Orion.\label{fig:orion1}}
\end{figure}

\subsection{Semantics: Schema and Data Update}

\begin{figure}[!ht]
  \centering
  \includegraphics[width=.48\textwidth]{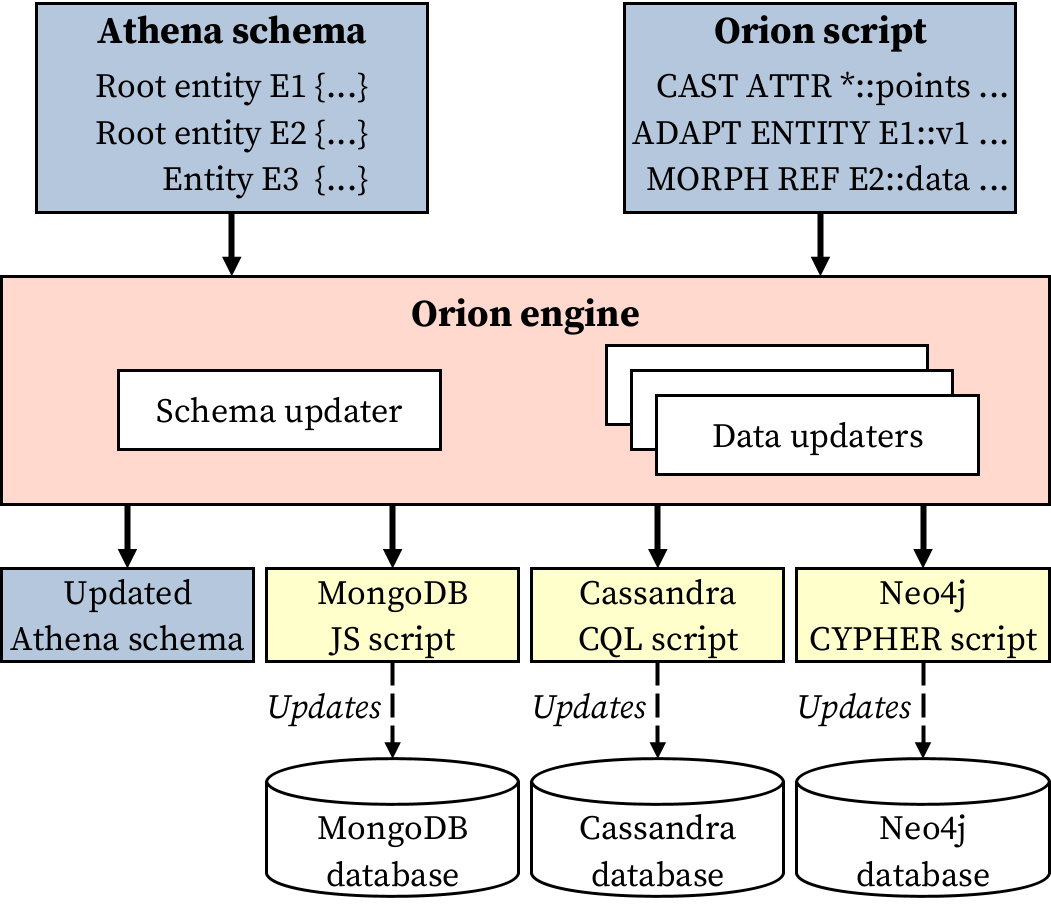}
  \caption{The Orion engine developed to handle schema and data updates.\label{fig:oriongen}}
\end{figure}

The Orion semantics is determined by the changes that each operation causes
in the existing schema and stored data. In our case, this semantics is
implemented by the Orion engine that updates the \uschema{} model and
translates Orion scripts into database-specific operations for updating
data according to the modified schema. The Orion engine has a component for
each of these two tasks: The \emph{Schema Updater} and the \emph{Data
  Updater}, as shown in Figure~\ref{fig:oriongen}.

The \emph{Schema Updater} takes an Athena model (i.e.,~a schema) and an
Orion model as input, and outputs the updated schema. To do so, the input
schema is taken as a starting point, and each Orion operation is
sequentially applied to that schema, with the semantics shown in
Section~\ref{sec:taxonomy}. This process is implemented as a
\emph{model~to~model} transformation which assures platform-independence.
It also has to be executed along with the \emph{Data Updater}, to assure
the correctness of the data adapting, or can be executed as a standalone
process to study schema evolution.

While we implemented a single \emph{Schema Updater}, as it is independent
of any database and works at a logical level, the \emph{Data Updater} is
bound to a specific database, so a different data updater must be
implemented for each supported system. In our case, we have developed data
updaters for MongoDB, Cassandra, and Neo4j. With this choice, we covered
document, columnar, and graph NoSQL data models, and we support three
widely used NoSQL stores, which occupy the position~5,~11, and~20 in the
DB-engines ranking as of March~2022.

A data updater receives an Athena and Orion model as input, and generates
the piece of database-specific code that applies the necessary changes on
the database according to the operations specified in the Orion script.
Therefore, this process consists of a \emph{model~to~text} transformation.

In the case of MongoDB, the data updater generates native MongoDB commands
and stores them in a Javascript file. Since MongoDB does not have to
declare an explicit schema, in order to apply changes, documents belonging
to the desired entity type have to be selected. To improve performance,
Orion analyzes the scripts and optimizes operations that can be applied
sequentially on the same entity type, stacking them together into a single
\emph{bulk write}. Some complex operations, however, do not allow that
optimization, and they must be executed in their own \emph{aggregation
  pipeline}. An example of this can be seen in
Figure~\ref{fig:dataMongodb}, where operations \texttt{RENAME} and
\texttt{CAST} are applied on the same entity type and therefore can be
stacked.

\begin{figure}[!ht]
\begin{lstlisting}[language=mongodb]
Sales_department.SaleSummary.bulkWrite([
 // RENAME SaleSummary::completedAt TO isCompleted
 {updateMany: {
    filter: {},
    update: {$rename: {"completedAt": "isCompleted" }}}},
 // CAST ATTR SaleSummary::isCompleted TO Boolean
 {updateMany: {
    filter: {},
    update: [{$set: { "isCompleted": { $convert:
                    { input: "$isCompleted", to: 8 }}}}]}}
])
\end{lstlisting}%
\caption{Example of two operations stacked together in MongoDB.\label{fig:dataMongodb}}
\end{figure}

In Cassandra, CQL (\emph{Cassandra Query Language}) instructions are
generated to perform the data update. Due to Cassandra declaring an
explicit schema, evolution changes are restricted. To implement some of
these operations it is necessary to export the data to an external file,
change the schema and import the data back.

The Neo4j data updater uses the Cypher language to generate code for
updating data. Given its graph nature, Neo4j is also able to handle
relationship types and therefore allows the full set of schema type
operations to be implemented for relationships. The schemaless nature of
Neo4j also allows the database to be updated in a similar way to
MongoDB:~(i)~Selecting all nodes belonging to the entity type to be
modified and~(ii)~applying the desired change. This also allows to stack
together changes to the same schema type, reducing the overhead of the
change.

Table~\ref{tab:implementation} sums up how each data updater handles each
operation. For each specific database, some keywords give insight of how
each operation is implemented, and also which operations cannot be executed
in a particular database.


\section{Case Studies of Orion Applications\label{sec:casestudy}}

\subsection{Case Study 1: A StackOverflow Refactoring}

Database refactoring is an activity aimed to improve the database design
and performance without changing its semantics~\cite{refactoring-DB}. A
refactoring is a small change on the schema, and several refactorings can
be applied to achieve a determined improvement. In our first case study,
Orion was used to apply a refactoring to a Neo4j database that imported the
StackOverflow\footnote{\url{https://archive.org/details/stackexchange}.}
dataset.

\begin{figure}[!ht]
  \centering
  \includegraphics[width=.5\textwidth]{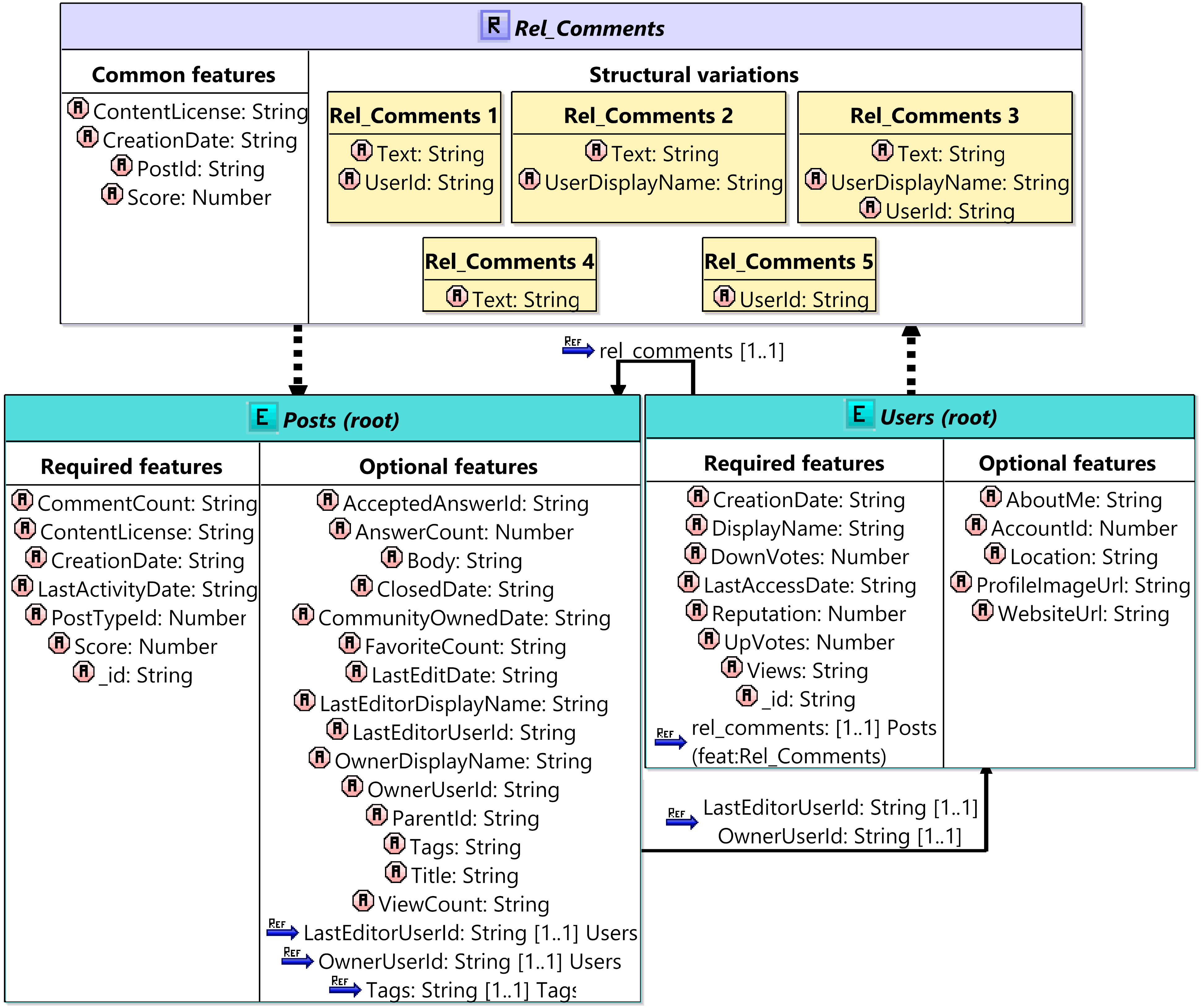}
  \caption{Excerpt of the {\tt StackOverflow} schema injected in Neo4j.\label{fig:sofschema}}
\end{figure}

We injected the dataset into Neo4j but changed it slightly during injection
to take advantage of relationship types. In StackOverflow, a
\texttt{Comment} references its \texttt{User} and a \texttt{Post} in
a~$1..1$ relation, so we injected this database transforming the
\texttt{Comment} entity type as a relationship type named
\texttt{Rel\_Comments} between \texttt{Users} and \texttt{Posts}. After
injecting the dataset into Neo4j, the schema inference strategy
from~\cite{metamodel2021} was applied. Figure~\ref{fig:sofschema} shows an
excerpt of the schema inferred with two of the seven entity types
discovered and the already mentioned new relationship type; the schema is
visualized with the notation introduced in~\cite{alberto-erforum2017}.
Here, \texttt{Posts} and \texttt{Users} are shown as \emph{union entity
  types} listing their required and optional features. \texttt{Users}
references \texttt{Posts} by \texttt{rel\_comments}, a reference with
attributes that belongs to \texttt{Rel\_Comments}. \texttt{Rel\_Comments}
has four required features (\texttt{ContentLicense}, \texttt{CreationDate},
\texttt{PostId} and \texttt{Score}) and five structural variations each
with a different set of additional features.

Analyzing the schema, we realized that the newly created relationship type
could be improved by casting some types of attributes to specific Neo4j
types, adding new fields or copying fields from \texttt{Users} and
\texttt{Posts} to the relationship type. Also injecting an entity type as a
relationship type caused that some attributes in \texttt{Rel\_Comments}
turned obsolete and could be deleted. In this way, by slightly changing the
schema, query performance could be improved.

The proposed refactoring can be divided into two blocks:~(i)~Applying
operations to some fields of \texttt{Users} and \texttt{Posts} to improve
query performance over them, and~(ii)~applying operations over the newly
created \texttt{Rel\_Comments} to improve its expressiveness.

The Orion script to refactor the StackOverflow schema is shown in
Figure~\ref{fig:soforion}. Firstly, some \emph{Cast} operations are
performed to convert certain fields stored as \emph{strings} to
\emph{timestamps}. These casts are performed against every schema type
containing the \texttt{CreationDate} and \texttt{LastAccessDate} (which are
all three schema types shown). Then a \emph{Mult} operation to allow the
possibility for a post to hold more than one tag and two \emph{Copy}
operations to move a couple of attributes from \texttt{Users} and
\texttt{Posts} to each \texttt{Comment} between them, in order to get quick
access to those fields. Operations regarding \texttt{Rel\_Comment} include
a \emph{Union} in order to maintain only a single variation and make all
the features mandatory, two \emph{Add Attribute} operations to create new
features, a \emph{Cast} over a feature that should be of \emph{double}
type, and two \emph{Delete} operations over the two \texttt{PostId} and
\texttt{UserId} carried from the injection that now are useless since the
relationship stores that information. Finally, we performed a \emph{Rename
  Relationship} to change the \texttt{Rel\_Comments} name to a more
suitable \texttt{comments} name for a relationship. Given this script and
the extracted schema, the Orion engine generates the updated schema and the
Neo4j API code script to execute the changes on the data.

\begin{figure}[!ht]
\begin{lstlisting}[language=orion]
StackOverflow_ops operations

Using stackoverflow:1

CAST ATTR *::CreationDate, LastAccessDate TO Timestamp
MULT REF Posts::Tags TO +
COPY Posts::PostTypeId TO Rel_Comments::CommentTypeId
                       WHERE id=PostId
COPY Users::Reputation TO Rel_Comments::UserReputation
                       WHERE id=UserId

UNION RELATIONSHIP Rel_Comments

ADD ATTR  Rel_Comments::LastEditDate: Timestamp
ADD ATTR  Rel_Comments::KarmaCount:   Number
CAST ATTR Rel_Comments::Score      TO Double
DELETE Rel_Comments::PostId, UserId

RENAME RELATIONSHIP Rel_Comments TO comments
\end{lstlisting}
  \caption{Operations to be applied to the \texttt{StackOverflow} schema and data.\label{fig:soforion}}
\end{figure}

\subsection{Case Study 2: Outlier Migration in Reddit}

\begin{figure}[!ht]
  \centering
  \includegraphics[width=.5\textwidth]{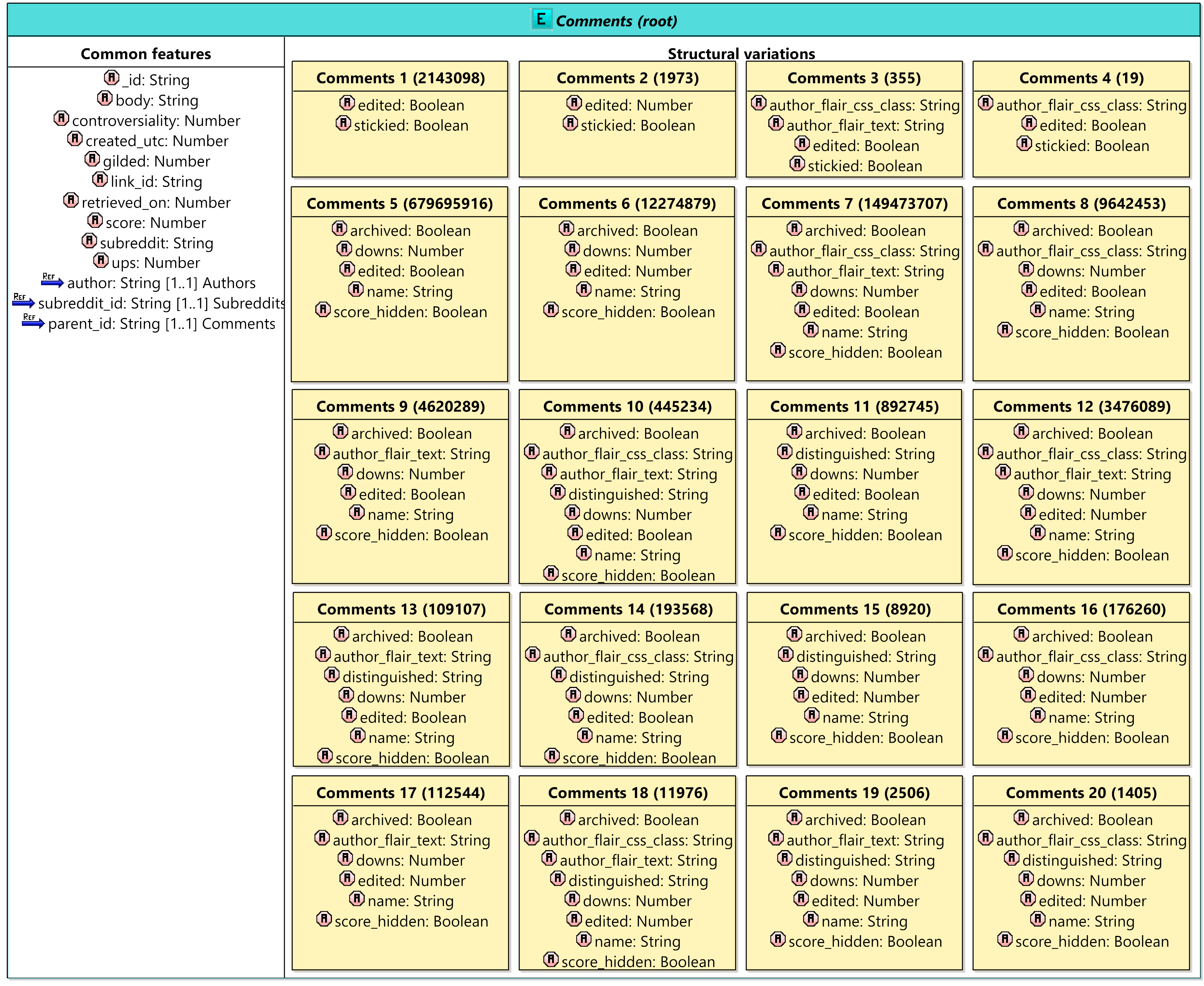}
  \caption{The \texttt{Comments} entity type from the \emph{Reddit} schema.\label{fig:reddit}}
\end{figure}

In this second case study, we remove outliers in the Reddit
dataset\footnote{\url{https://files.pushshift.io/reddit/comments/}.}. We
injected this dataset into MongoDB, and inferred its schema by applying the
process in~\cite{metamodel2021}. Figure~\ref{fig:reddit} shows the inferred
\texttt{Comment} entity type, with more than~860 million comments
distributed in~20 structural variations. We consider that outliers are
those variation with a very small number of objects.

\begin{figure}[!ht]
  \centering
  \includegraphics[width=.5\textwidth]{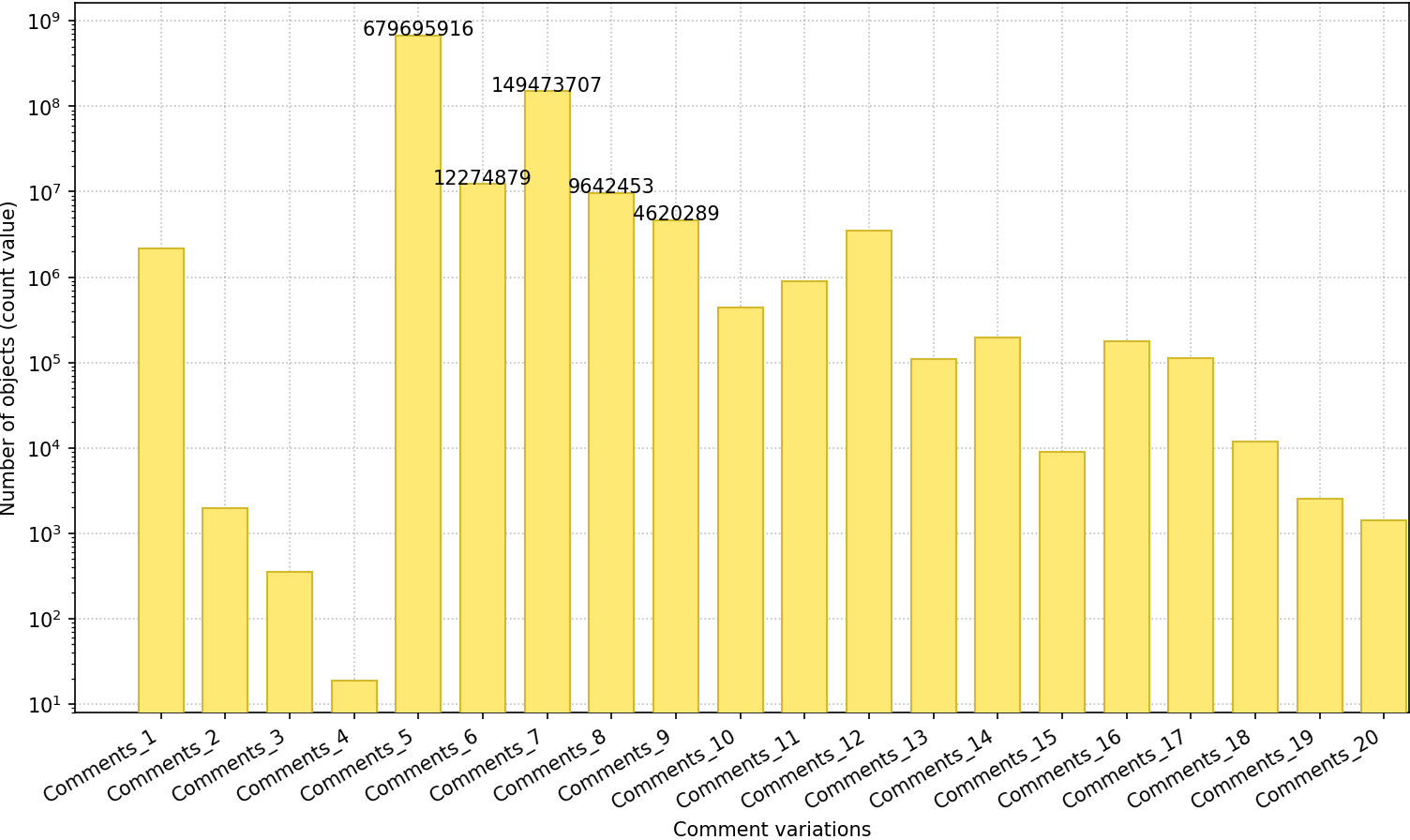}
  \caption{\texttt{Comments} variations represented by their \emph{count}
    property.\label{fig:outliers}}
\end{figure}

We developed a process in which outliers could be detected by using the
\emph{count} property of the variations, that indicates how many instances
belong to each variation, and then we proposed an approach to migrate those
outlier instances to regular variations (i.e.,~variations that were not
outliers). In Figure~\ref{fig:outliers} we show a bar chart with
logarithmic axes in which each variation is represented by its \emph{count}
property. There, few variations hold the majority of objects. By following
our approach, we classified the top five most populated variations as
\emph{regular}, and the other fifteen variations as \emph{outliers}.

Using Orion, a developer can delete obsolete instances belonging to old
variations, as well as adapt some variations to new ones, migrating the
data accordingly. The developer can do so by using the \emph{Delvar} and
\emph{Adapt} operations, as is shown in Figure~\ref{fig:outlier_script}.
Here, adapting variation~11 (an outlier) to variation~5 (a regular
variation) means that all instances matching variation~11 will be modified
accordingly to fit variation~5, reducing the number of resulting
variations.

\begin{figure}[!ht]
\begin{lstlisting}[language=orion]
Reddit_migration operations

Using reddit:1

DELVAR ENTITY Comments::v1
DELVAR ENTITY Comments::v2
DELVAR ENTITY Comments::v3
DELVAR ENTITY Comments::v4

ADAPT ENTITY Comments::v10 TO v7
ADAPT ENTITY Comments::v11 TO v5
ADAPT ENTITY Comments::v12 TO v7
ADAPT ENTITY Comments::v13 TO v9
ADAPT ENTITY Comments::v14 TO v8
ADAPT ENTITY Comments::v15 TO v6
ADAPT ENTITY Comments::v16 TO v6
ADAPT ENTITY Comments::v17 TO v6
ADAPT ENTITY Comments::v18 TO v7
ADAPT ENTITY Comments::v19 TO v6
ADAPT ENTITY Comments::v20 TO v6
\end{lstlisting}%
\caption{Orion script used to migrate variations on Reddit \texttt{Comments}.\label{fig:outlier_script}}
\end{figure}

Each of these \emph{Delvar} and \emph{Adapt} operations will be translated
into Javascript code and will remove or migrate instances from a certain
variation. In Figure~\ref{fig:reddit_adapt} an example of the code
generated from one of the \emph{Adapt} operations is shown, where a match
that captures only instances of variation~11 is applied and then fields are
removed and added with default values as needed. Once the script is
executed against the database, data is migrated, variations are removed and
the complexity of the schema is reduced as a result.

\begin{figure}[!ht]
\begin{lstlisting}[language=mongodb]
// ADAPT ENTITY Comments::11 TO 5
reddit.Comments.updateMany({
  "archived":               {$exists: true},
  "distinguished":          {$exists: true},
  "downs":                  {$exists: true},
  "edited":                 {$exists: true},
  "name":                   {$exists: true},
  "score_hidden":           {$exists: true},
  "author_flair_css_class": {$exists: false},
  "author_flair_text":      {$exists: false}},
  [
    {$unset: ["distinguished"]}
  ])
\end{lstlisting}%
\caption{Orion script migrating \texttt{Comments} variation~11 to~5.\label{fig:reddit_adapt}}
\end{figure}

\section{Evaluation\label{sec:evaluation}}

Two kinds of evaluations have been carried out. The schema change semantics
of each operation of the proposed taxonomy has been formally validated by
using Alloy. Also, the feasibility or applicability of the changes has been
evaluated by measuring execution times for the three currently supported
NoSQL systems.

\subsection{Validating the Taxonomy \label{sec:alloy}}

Alloy~5\footnote{\url{https://alloytools.org/}.} was used to implement each
schema change based on its pre and postconditions. This has been achieved
by applying a three step process in which~(i)~\uschema{} concepts and their
restrictions have been modeled,~(ii)~operations implementing the taxonomy
have been defined, and then~(iii)~\emph{checks} for contradictions have
been implemented for each operation. Each step will be detailed below.

The \uschema{} metamodel has been modeled in Alloy by using
\emph{signatures}. In~\ref{fig:alloy_uschema} an excerpt of \uschema{} is
shown, consisting of two parts:~(i)~\emph{entities} and
\emph{relationships} field declarations, which are a set of \emph{Entity
  types} and \emph{Relationship types}, and~(ii)~a set of facts
implementing restrictions that any \uschema{} model must fulfill, such
as:~(i)~A schema must contain at least one entity type or one relationship
type,~(ii)~there cannot be two different entity types with the same name,
and~(iii)~each reference to a schema type must belong to the same schema as
that schema type. Once the \uschema{} specification is defined, Alloy is
capable of searching for scenarios that fulfill all the provided
restrictions.

\begin{figure}[!ht]
\begin{lstlisting}[language=alloy]
some sig USchema
{
 entities: set EntityType,
 relationships: set RelationshipType
}
{
 some entities or some relationships
 entities.parents in entities
 relationships.parents in relationships

 some e: entities          | e.root = True
 all e1, e2: entities      | e1.name = e2.name => e1 = e2
 all r1, r2: relationships | r1.name = r2.name => r1 = r2
 all ref: entities.variations.features
   + relationships.variations.features
                 | ref.refsTo in entities
 all aggr: entities.variations.features
    + relationships.variations.features
                 | aggr.aggregates in entities.variations
}
\end{lstlisting}
  \caption{The \uschema{} definition excerpt in
    Alloy.\label{fig:alloy_uschema}}
\end{figure}

The next step is to model the change operations in the taxonomy as Alloy
operations, by using \emph{predicates} that may be applied over instances
of \uschema{} elements. Each operation shows the same structure:~(i)~it
checks that input parameters do meet the preconditions, and then~(ii)~it
matches the changes to be reflected on the output parameters.

\begin{figure}[!ht]
\begin{lstlisting}[language=alloy,escapeinside={(*}{*)}]
pred Operation_RenameEntity [schemaI, schemaO: USchema,
                             entityI, entityO: EntityType,
                             newName: SchemaTypeName]
{
 entityI in schemaI.entities and
 entityO not in schemaI.entities
 // Precondition check: (*$n \notin T.names$*)
 newName not in schemaI.entities.name

 entityO.name       = newName
 entityO.root       = entityI.root
 entityO.parents    = entityI.parents
 entityO.variations = entityI.variations
 schemaO.entities   = schemaI.entities - entityI + entityO
 schemaO.relationships = schemaI.relationships
}
\end{lstlisting}
\caption{Alloy definition for the \emph{Rename Entity} operation.\label{fig:alloy_rename}}
\end{figure}

\begin{figure}[!ht]
  \centering
  \includegraphics[width=.5\textwidth]{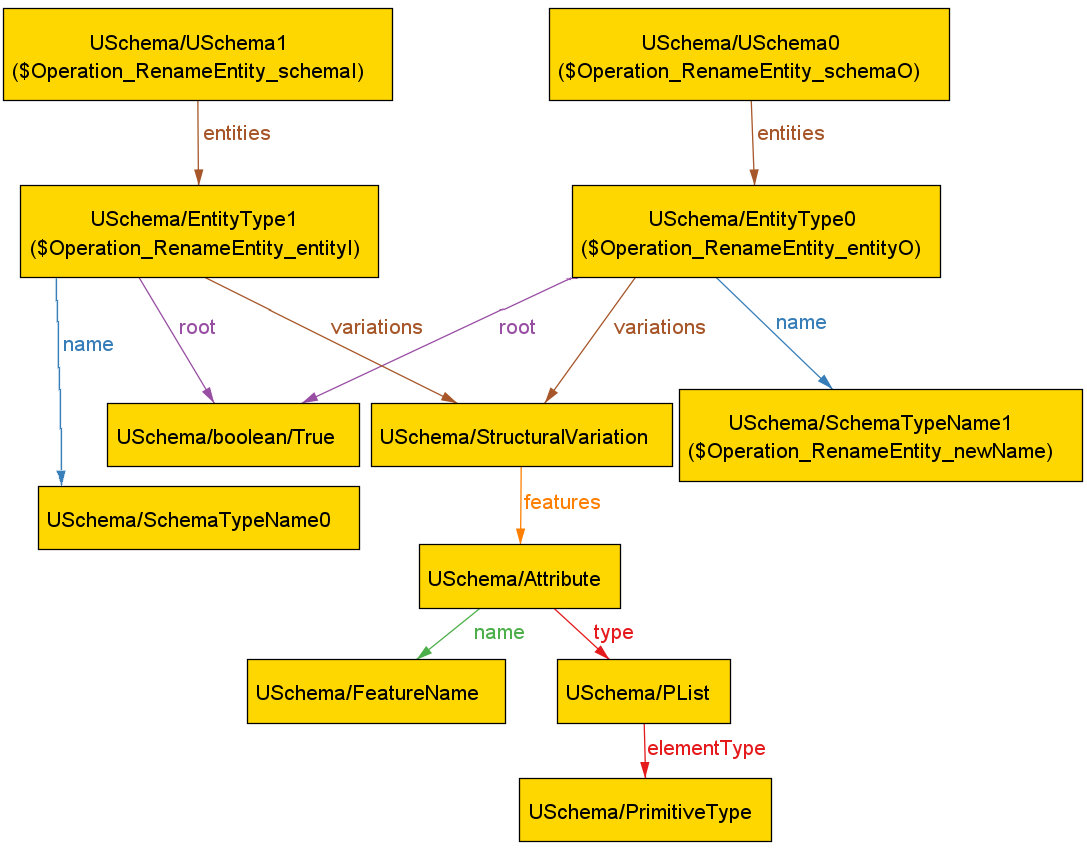}
  \caption{One of the \emph{Rename Entity} scenarios found.\label{fig:alloy_scenario}}
\end{figure}

In Figure~\ref{fig:alloy_rename}, the \emph{Rename Entity} operation is
implemented. Its precondition is declared in the same way as it was defined
in Table~\ref{tab:taxonomy}, \texttt{newName not in schemaI.entities.name},
and then several statements are defined to be fulfilled by the output
schema. When this operation is executed in Alloy to search scenarios in
which is sucessfully applied, a scenario is found, which is shown in
Figure~\ref{fig:alloy_scenario}. As can be seen, the input schema only has
an entity type whose \emph{name} changes in the output schema but its
\emph{root} and \emph{variations} properties are the same.

We have defined Alloy \emph{Check} operations to find contradictions, for
instance, check operation for the \emph{Rename Entity} operation is shown
in Figure~\ref{fig:alloy_check}. When executing each check operation, no
scenarios were found in which the implications (i.e.,~postconditions) of
the operation are not true (counterexample).

\begin{figure}[!ht]
\begin{lstlisting}[language=alloy,escapeinside={(*}{*)}]
Check_Operation_RenameEntity: check
{
 all schemaI, schemaO: USchema,
     entityI, entityO: EntityType,
     newName: SchemaTypeName |
  Operation_RenameEntity[schemaI, schemaO,
                         entityI, entityO,
                         newName] =>
    // Postcondition check: (*$t.name = n$*)
    entityO.name = newName
    // Invariant check: Everything else remains the same.
    and CheckSchemaEquality[schemaI.entities - entityI,
                            schemaO.entities - entityO,
                            schemaI.relationships,
                            schemaO.relationships]
} for 10
\end{lstlisting}
\caption{Postcondition checking of the \emph{Rename Entity} operation.\label{fig:alloy_check}}
\end{figure}

Therefore, we concluded that preconditions were consistent and
postconditions were valid. The usage of Alloy also served to refine with
additional preconditions certain operations, such as
\emph{Extract/Split/Merge~Entity}, which were not consistent at the
beginning of the process. It also showed the importance of including
invariants in the metamodel.


\subsection{Measuring performance of Orion operations\label{ssec:performance}}

To evaluate the feasibility of each implemented operation we created the
following scenario for each database system considered. First, we defined a
schema with several root entity types:~one entity type per group of
operations (features, attributes, references, and aggregates), and one
entity type per schema type operation, each one of them with the same
number of features. Then by using the tool shown
in~\cite{alberto-comonos2020}, we generated a dataset of~150,000 instances
per entity type conforming to that schema. After that, we defined a process
to inject this dataset into a MongoDB, Cassandra, and Neo4j. Then we
formulated an experiment, but first we applied several standard queries to
warm up the database and let it fill its caches. The designed warm up
queries do scan the entire database looking for non-existing values on
non-indexed fields.

In order to provide a meaningful expression of the feasibility of the
implemented SCOs, we did not measure absolute times. Instead, we used a
\emph{modification operation} $op_{mod}$ to normalize the obtained times.
This $op_{mod}$ operation modifies a field that is not indexed, so the
database is not optimized for it and results are more reliable. In the
third step, $op_{mod}$ is applied over all the instances of a certain
entity type. Note that an \emph{update} operation is preferred over a
standard query because we are measuring operations modifying the database.

The final step consists on executing each operation independently to
measure its execution time. To do so, we defined three blocks to
execute:~(i)~Entity type operations,~(ii)~feature, attribute, reference,
and aggregate operations, and~(iii)~relationship type operations, if
applicable. Since operations are executed individually, it is not possible
to take advantage of certain mechanisms such as stacking operations
together, which is relevant in the case of, for example, MongoDB. This
whole process was repeated five times to get a reliable mean time. Given
the different nature of each database system considered, the $op_{mod}$
operation is slightly different for each one of them.

Table~\ref{tab:implementation} shows the different execution times for each
taxonomy operation performed over each of the considered NoSQL systems. The
table includes two columns for each system, one shows a summary of the
native code performed and the other the execution time multiplied by a
factor that is the execution time for the $op_{mod}$ operation on MongoDB,
Cassandra and Neo4J, which are denoted as $t_M$, $t_C$ and $t_N$,
respectively.

MongoDB operations performed as expected because the majority of them scan
over a single entity type (\emph{Delete}, \emph{Unnest} or \emph{Cast}), so
their ratio is close to~$1\times t_M$ and only a couple of operations such
as \emph{Copy} or \emph{Morph} do require additional scans (or an explicit
\emph{join}) and therefore are much more costly. As was explained in
Section~\ref{sec:taxonomy}, although \emph{Delvar} and \emph{Adapt} are
semantically equal, they are implemented differently because the former
removes instances belonging to a variation and the latter transforms those
instances to a new variation by adding and/or deleting fields.

Cassandra operations do not show huge performance differences between them,
although the ones with the \texttt{COPY} command are the most costly. As
explained in Section~\ref{sec:orion}, these operations are the ones that
were implemented by means of an export/import to an external file. These
tables were of only five fields, but it is foreseeable that their
performance would drop if tables had more fields. It is also important to
note that CSV manipulation on the most costly operations was not included
in the measurement.

Finally, Neo4j operations behaved in a similar way as in MongoDB, although
certain relationship operations (\emph{Split}, \emph{Merge}, and
\emph{Union}) performed worse than other relationship operations because
they not only affect single relationships but also involve creating new
relationships between nodes and filling their fields.

\begin{table*}[!t]
{
\caption{Implementation and performance of operations.\label{tab:implementation}}
\begin{adjustbox}{width=1\textwidth}
\tiny\sffamily%
\noindent%
\begin{tabular}{p{.05\textwidth}p{.28\textwidth}p{.02\textwidth}p{.28\textwidth}p{.02\textwidth}p{.28\textwidth}p{.02\textwidth}}
  \toprule
  & \multicolumn{1}{c}{\textbf{MongoDB}} & $\mathbf{t_M}$ & \multicolumn{1}{c}{\textbf{Cassandra}} & $\mathbf{t_C}$ & \multicolumn{1}{c}{\textbf{Neo4j}} & $\mathbf{t_N}$\\
\midrule
\multicolumn{2}{l}{\textbf{Entity type Operations}}\\
\midrule
{\bfseries Add}     & createCollection(),\$addFields & $0.01$ & CREATE \emph{table} & $0.21$ & CREATE,MATCH,SET \emph{node} & $0.37$ \\
{\bfseries Delete}  & drop()                         & $0.01$ & DROP \emph{table} & $0.37$ & MATCH,DELETE \emph{node}    & $1.12$ \\
{\bfseries Rename}  & renameCollection()             & $0$    & 2$\times$(COPY,DROP,CREATE) \emph{table} & $2.86$ & MATCH,REMOVE,SET \emph{node} & $1.89$ \\
{\bfseries Extract} & \$project,\$out                & $0.31$ & 2$\times$COPY \emph{table},CREATE \emph{table} & $2.25$ & MATCH,CREATE \emph{node} & $2.92$ \\
{\bfseries Split}   & 2$\times$(\$project,\$out),drop()     & $0.63$ & (4$\times$COPY,2$\times$CREATE,DROP) \emph{table} & $4.44$ & MATCH,2$\times$CREATE,DELETE \emph{node} & $6.55$ \\
{\bfseries Merge}   & 2$\times$\$merge,2$\times$drop()      & $4.95$ & (4$\times$COPY,CREATE,2$\times$DROP) \emph{table} & $4.72$ & 2$\times$MATCH,CREATE,2$\times$DELETE \emph{node} & $8.29$ \\
{\bfseries Delvar}  & remove()                       & $0.38$ & --- & --- & MATCH,DELETE \emph{node} & $1.42$ \\
{\bfseries Adapt}   & \$unset,\$addFields            & $0.70$ & --- & --- & MATCH,REMOVE,SET \emph{node} & $1.95$ \\
{\bfseries Union}   & \$addFields                    & $1.40$ & --- & --- & MATCH,SET \emph{node} & $8.44$ \\
\midrule
\multicolumn{2}{l}{\textbf{Relationship type Operations}}\\
\midrule
{\bfseries Add}     & --- & --- & --- & --- & --- & --- \\
{\bfseries Delete}  & --- & --- & --- & --- & MATCH,DELETE \emph{rel} & $2.43$ \\
{\bfseries Rename}  & --- & --- & --- & --- & MATCH,apoc.refactor.setType \emph{rel} & $0.22$ \\
{\bfseries Extract} & --- & --- & --- & --- & MATCH,CREATE \emph{rel} & $5.43$ \\
{\bfseries Split}   & --- & --- & --- & --- & MATCH,2$\times$CREATE \emph{rel} & $10.99$ \\
{\bfseries Merge}   & --- & --- & --- & --- & 2$\times$MATCH,CREATE,2$\times$DELETE \emph{rel} & $13.84$ \\
{\bfseries Delvar}  & --- & --- & --- & --- & MATCH,DELETE \emph{rel} & $0.84$ \\
{\bfseries Adapt}   & --- & --- & --- & --- & MATCH,REMOVE,SET \emph{rel} & $0.76$ \\
{\bfseries Union}   & --- & --- & --- & --- & MATCH,SET \emph{rel} & $15.93$ \\
\midrule
\multicolumn{2}{l}{\textbf{Feature Operations}}\\
\midrule
{\bfseries Delete}  & \$unset  & $1.08$ & DROP \emph{column} & $0.22$ & MATCH,REMOVE \emph{field} & $0.78$ \\
{\bfseries Rename}  & \$rename & $1.22$ & 2$\times$COPY \emph{table},DROP \emph{column},ADD \emph{column} & $2.07$ & MATCH,SET,REMOVE \emph{field} & $2.03$ \\
{\bfseries Copy}    & \$lookup,\$addFields,\$addFields,\$out & $4.06$ & 2$\times$COPY \emph{table},ADD \emph{column} & $2.21$ & 2$\times$MATCH,SET \emph{field} & $2.24$ \\
{\bfseries Move}    & \$lookup,\$addFields,\$addFields,\$out,\$unset & $5.09$ & 2$\times$COPY \emph{table},ADD \emph{column},DROP \emph{column} & $2.30$ & 2$\times$MATCH,SET,REMOVE \emph{field} & $3.31$\\
{\bfseries Nest}    & \$rename & $1.27$ & --- & --- & --- & --- \\
{\bfseries Unnest}  & \$rename & $1.30$ & --- & --- & --- & --- \\
\midrule
\multicolumn{2}{l}{\textbf{Attribute Operations}}\\
\midrule
{\bfseries Add}     & \$addFields & $1.35$ & ADD \emph{column} & $0.21$ & MATCH,SET \emph{field} & $0.76$ \\
{\bfseries Cast}    & \$set       & $1.31$ & 2$\times$(COPY,DROP,CREATE) \emph{table} & $3.06$ & MATCH,SET \emph{field} & $1.50$ \\
{\bfseries Promote} & ---         & ---    & 2$\times$(COPY,DROP,CREATE) \emph{table} & $3.08$ & CREATE \emph{constraint} UNIQUE & $4.12$ \\
{\bfseries Demote}  & ---         & ---    & 2$\times$(COPY,DROP,CREATE) \emph{table} & $3.08$ & DROP \emph{constraint} & $0.03$ \\
\midrule
\multicolumn{2}{l}{\textbf{Reference Operations}}\\
\midrule
{\bfseries Add}     & \$lookup,\$addFields,\$out & $4.09$ & ADD \emph{column},2$\times$COPY \emph{table} & $2.04$ & 2$\times$MATCH,CREATE \emph{rel} & $5.39$ \\
{\bfseries Cast}    & \$set       & $1.46$ & 2$\times$(COPY,DROP,CREATE) \emph{table} & $3.07$ & --- & --- \\
{\bfseries Mult}    & \$set       & $1.41$ & --- & --- & --- & --- \\
{\bfseries Morph}   & \$lookup,\$addFields,\$out,\$unset & $4.95$ & --- & --- & --- & --- \\
\midrule
\multicolumn{2}{l}{\textbf{Aggregate Operations}}\\
\midrule
{\bfseries Add}     & \$addFields     & $1.43$  & CREATE \emph{type},ADD \emph{column} & $0.24$ & --- & --- \\
{\bfseries Mult}    & \$set           & $1.45$  & --- & --- & --- & --- \\
{\bfseries Morph}   & insert(),save() & $34.08$ & --- & --- & --- & --- \\
\bottomrule
\end{tabular}
\end{adjustbox}
}
\end{table*}


\section{Related Work\label{sec:relatedwork}}

In this section, we will contrast our proposal to some relevant schema
evolution approaches presented for relational and object-oriented
databases, and to most of research work done on NoSQL systems.

The works of Jean Luc Hainaut et al.~\cite{hick2003, hainaut2006} and Carlo
Curino et al.~\cite{curino-prism2008,curino-evolution2013} are some of more
influential contributions on automating relational schema evolution. While
the interest of Curino et al. was exclusively focused on relational systems
in order to build the PRISM/PRISM++ tool, Hainaut et al. defined the
DB-MAIN generic approach that involved the main data models existing at the
end of nineties.

DB-Main was based on two main elements:~(i)~The Generic Entity/Relationship
(GER) metamodel to achieve platform-independence; and~(ii)~a
transformational approach to implement processes such as reverse and
forward engineering, and schema mappings. Our proposal is also based on a
generic metamodel and a transformational approach, but differs in two
several significant aspects. Firstly, GER did not integrate data models
supported by NoSQL systems, instead we used the \uschema{} metamodel which
was specially designed to support NoSQL and relational schemas. Secondly,
we have taken advantage of Model-driven Engineering (MDE) technology
incorporated in the EMF/Eclipse framework, as \uschema{} data model is
implemented in form of an Ecore metamodel~\cite{steinberg-emf2009}. A
detailed comparison between GER and \uschema{} data model is given
in~\cite{metamodel2021}. Also, it should be noted that no taxonomy was
defined for DB-Main. Instead, the taxonomy shown
in~\cite{roddick-taxonomy1993} is adopted.

The PRISM/PRISM++ tool is aimed at automating data migration tasks and
rewriting legacy queries. PRISM/PRISM++ provides an evolution language
based on \emph{Schema Modification Operators} (SMOs) that preserve
information and are revertible, and \emph{Integrity Constraint Modification
  Operators} (ICMO). Given a schema, a new schema, and a set of mappings
expressed through SMOs and ICMOs, queries can be rewritten and stored data
are updated. Although much more mature and evolved than our work, this
approach does not address the NoSQL database evolution.

In OO systems, schema evolution is a more complicated problem than in
relational systems. This is because OO schemas are classes hierarchies of
inheritance and aggregation, while relational schemas are sets of tables.
In addition, classes have structure (attributes) and behavior (methods). OO
schema evolution aroused great interest until the mid-1990s, when OO
systems evidenced limitations to become an alternative to relational
systems. A survey on that topic was presented in~\cite{roddick92}, and
Banerjee et al.~\cite{banerjee-ooevolution1987} published a seminal paper
proposing a schema changes taxonomy, and discussing the operations whose
semantic impact was analyzed. Our proposal is inspired by that work: we
have defined a taxonomy for NoSQL databases, the change operations are
rigorously specified and its performance is measured.

To the best of our knowledge, most research efforts on NoSQL schema
evolution are considered below. A proposal for different NoSQL databases is
described in~\cite{klettke-bigdata2016}. The main focus in this approach is
on efficient data migration for different NoSQL databases. In this work,
a~5-operation taxonomy is defined for a simple data model: schemas are a
set of entities that are formed by attributes whose type can be a primitive
or collection type or either another entity, but relationships between
entities are not considered. In this work, the schema model serves as an
abstraction layer on top of different NoSQL databases but do provide the
additional constraints of a conceptual model. The five schema evolution
operations are \emph{add/remove/rename} properties, and \emph{copy/move} a
set of properties from an entity type to another. This taxonomy was
implemented in Darwin~\cite{Darwin:ICDE, storl-darwin2022}, a data platform
for schema evolution management and data migration. Darwin is also able to
extract the implicit schema and version history of NoSQL databases, manage
those versions, update data eager, lazily or with intelligent hybrid
approaches and rewrite queries that try to retrieve data that is yet to be
updated. It was also implemented on the Google Cloud Platform as part of
the Cleager tool~\cite{scherzinger-cleager2015}. This tool maps operations
of the taxonomy to MapReduce functions. Our proposal is based on a more
complex unified data model, and this results in a richer change taxonomy,
which is applicable for NoSQL and relational data models.

As noted in~\cite{ram1991}, heterogeneous database systems are commonly
implemented through a unified schema approach or a multi-database approach.
Holubová et al. explored schema evolution for multi-database database
systems in~\cite{holubova-evolution2021} and~\cite{vavrek-mmevolver2019}.
They proposed a layered architecture which consists of a model-independent
layer and a model-specific layer. The former delegates to the corresponding
model-specific components by examining the prefix of the affected entities,
thus providing a way to support both intra-model and inter-model
operations. Since heterogeneous databases can store entities referencing
others stored in a different data model, the proposal also provides the
foundation for managing referential integrity between them when modifying
the database. Regarding schema evolution, a taxonomy of~10 operations is
defined:~5 for entity types (\emph{kinds}) and~5 for properties. These
latter correspond to those defined in~\cite{klettke-bigdata2016}, and the
first~5 are \emph{add}, \emph{drop}, \emph{rename}, \emph{split}, and
\emph{merge}, which have the same meaning as in our taxonomy. The impact of
operations is discussed classifying them as intra-model or inter-model,
depending on how many models are affected by changes, and as global and
local operations, depending on whether they may be specified over the
global union schema, or only over a specific model. It is worth noting that
we have not tackled the issues related to schema evolution in heterogeneous
systems, but we are interested in offering automation for individual stores
in a data model-independent way through a unified data model. Furthermore,
our taxonomy includes operations related to relationships and variations,
and we have defined a complete language to define and execute schema change
operations.

In~\cite{cleve-typhon2020}, a taxonomy is proposed as part of an approach
to rewrite queries for polystore (i.e.,~heterogeneous databases) evolution.
The taxonomy includes six operations applicable to \emph{entity types},
four to \emph{attributes} and four to \emph{relations}. A generic language,
called TyphonML, is used to define relational and NoSQL schemas, physical
mapping and schema evolution operations. Like our approach, TyphonML is
based on a generic metamodel also created with the Ecore metamodeling
language. However, \uschema{} is a richer data model as discussed
in~\cite{metamodel2021}, which allowed us to define operations
on~(i)~aggregates and references in a separate way,~(ii)~structural
variations, and~(iii)~distinguish between entity and relationship types in
graph stores.

Suárez-Otero et al.~\cite{otero-evolution2020} have recently published a
work-in-progress paper where they define a taxonomy of~7 schema changes and
analyze how they affect to the schema and data in the case of Cassandra,
but no automation is addressed. In our case, a unified model for logical
schemas is considered, the taxonomy includes a larger number of operations,
and the taxonomy is implemented for~3 popular NoSQL systems for different
data models.

Some works have presented approaches for a particular NoSQL store. Loup
Meurice and Anthony Cleve presented a strategy for
MongoDB~\cite{meurice2017}: queries are extracted from Java code, and
queries are analyzed to find the database entities (i.e.,~collections of
documents), entity properties and references between documents. A
``historical database schema'' is obtained from the set of variations of
each collection, which is visualized through a table by using different
colors and icons to indicate if a property can give rise to errors due to
possible data corruption or warn developers of renamed properties or
collections. As far as we know, this work is the only proposal that
extracts schemas from code. The strategy of schema evolution is limited to
consider existing entity variations as schema changes. We proposed a
strategy for NoSQL and relational stores, and a operations definition
language. Actually, the work of Meurice and Cleve is a reverse engineering
strategy to extract schemas rather than a schema evolution approach.

KVolve~\cite{saur-evolving2016} is a library that allows for schema
evolution in the Redis\footnote{\url{https://redis.io}.} key-value store.
It is restricted to key and value changes for entries sharing a common
prefix, and accepts a previously-defined user function written in C with
the actions to be performed. In this library, key changes must be done by
unambiguous bijections, and value changes can only access the value to be
updated. This library to operate on standalone Redis instances. It employs
a lazy strategy that updates entries as they are accessed. This solution is
limited to Redis, while our approach is applicable to the four most widely
used NoSQL data models. In~\cite{metamodel2021}, a mapping of Redis to
\uschema{} is shown, so the approach here presented may be applied to
Redis, but we have not built an Orion engine yet.

In short, the differences between our work and the existing ones can be
summarized as follows. We suppose a NoSQL schema represented as a
\uschema{} model has been extracted from a existing store, this schema can
then be changed by writing a Orion script, and the schema and data updates
are automatically performed. Orion is a system-independent operation
language because \uschema{} is a unified data model that includes all the
typical elements of logical NoSQL and relational schemas, even structural
variations are considered, which allows a more complete taxonomy to be
defined.

Table~\ref{tab:related} summarizes the comparison carried out between our
approach and other works. Several criteria are defined to compare the
schema evolution approaches discussed above: changes operations in the
taxonomies, supported database paradigms, schema representation, aim, if
operation impact analysis has been performed, and if a tool is available.

\begin{sidewaystable*}[htb!]
{
 \caption{Comparison of schema evolution approaches\label{tab:related}}
 \tiny\sffamily
 \begin{tabular}{ccllllll}
 \toprule
  & & \textbf{Curino et al.}~\cite{curino-prism2008} &
      \textbf{Hainaut et al.}~\cite{hainaut1994} &
      \textbf{St{\"o}rl et al.}~\cite{klettke-bigdata2016} &
      \textbf{Holubová et al.}~\cite{holubova-evolution2021} &
      \textbf{Fink et al.}~\cite{cleve-typhon2020} &
      \textbf{Hernández et al.}\\
\midrule
{\multirow{6}{*}{\textbf{Operation}}} & \begin{tabular}[c]{@{}c@{}}\textbf{Schema}\\\textbf{Types}\end{tabular} &
  \begin{tabular}[l]{@{}l@{}}Create,Drop,Rename\\Copy,Merge,Join\\Partition,Decompose\end{tabular} &
  \begin{tabular}[l]{@{}l@{}}Add,Delete,Rename\\Change to/from weak\\Split,Partition,Join,Coalesce\end{tabular} &
  Create,Drop,Rename &
  \begin{tabular}[l]{@{}l@{}}Create,Drop,Rename\\Split,Merge\end{tabular} &
  \begin{tabular}{@{}l@{}}Add,Remove,Rename\\Merge,Split,Migrate\end{tabular} &
  \begin{tabular}[l]{@{}l@{}}Add,Delete,Rename\\Extract,Split,Merge\end{tabular}\\

\cmidrule{2-2} & \textbf{Variations} &
  --- &
  --- &
  --- &
  --- &
  --- &
  Delvar,Adapt,Union\\

\cmidrule{2-2} & \textbf{Features} &
  --- &
  --- &
  \begin{tabular}[l]{@{}l@{}}Add,Delete,Rename\\Copy,Move\end{tabular} &
  \begin{tabular}[l]{@{}l@{}}Add,Delete,Rename\\Copy,Move\end{tabular} &
  Add,Remove,Rename &
  \begin{tabular}[l]{@{}l@{}}Delete,Rename,Copy\\Move,Nest,Unnest\end{tabular}\\

\cmidrule{2-2} & \textbf{Attribute} &
  \begin{tabular}{@{}l@{}}Add,Drop,Rename\\Copy,Move\end{tabular} &
  \begin{tabular}{@{}l@{}}Add,Drop,Rename\\Type change\\Promote,Demote\end{tabular} &
  --- &
  --- &
  Type change &
  \begin{tabular}[l]{@{}l@{}}Add,Cast\\Promote,Demote\end{tabular}\\

\cmidrule{2-2} & \textbf{Reference} &
  --- &
  \begin{tabular}[l]{@{}l@{}}Add,Delete,Rename\\Cardinality change\end{tabular} &
  --- &
  --- &
  Cardinality change &
  Add,Cast,Mult,Morph\\

\cmidrule{2-2} & \textbf{Aggregate} &
  --- &
  --- &
  --- &
  --- &
  Cardinality change &
  Add,Mult,Morph\\

\cmidrule{1-2} \multicolumn{2}{c}{\textbf{Supported Paradigms}} &
  Relational &
  \begin{tabular}{@{}l@ {}}Relational, hierarchical\\Object-oriented, network\end{tabular} &
  NoSQL &
  Multi-model &
  Multi-model &
  Relational, NoSQL\\

\cmidrule{1-2} \multicolumn{2}{c}{\textbf{Unified Schema Representation}} &
  No &
  Yes (GER) &
  No (But a Generic Interface) &
  No &
  No &
  Yes (\uschema{})\\

\cmidrule{1-2} \multicolumn{2}{c}{\textbf{Data Update}} &
  Yes &
  Yes &
  Yes (eager, lazy, hybrid)&
  Yes &
  Yes &
  Yes \\

\cmidrule{1-2} \multicolumn{2}{c}{\textbf{Code Update}} &
  \begin{tabular}[l]{@{}l@{}}Query rewriting\\SQL Views\end{tabular} &
  Program modification (hints) &
  Query adaptation &
  Query adaptation &
  Query adaptation &
  No\\

\cmidrule{1-2} \multicolumn{2}{c}{\textbf{Implementing Tool}} &
  PRISM/PRISM++ &
  DB-Main &
  Darwin &
  MM-evolver &
  TyphonML &
  Orion\\
 \cmidrule{1-2} \multicolumn{2}{c}{\textbf{Semantic Change Analysis}} &
  No &
  No &
  Yes &
  Yes &
  Yes &
  Yes\\
 \bottomrule
 \end{tabular}%
}
\end{sidewaystable*}


\section{Conclusions and Future Work\label{sec:conclusions}}

In this paper we have explored the NoSQL schema evolution by using a
generic solution: a unified data model with which we have defined a
taxonomy of schema changes. We presented the Orion schema operation
language implementing this taxonomy. Thanks to the richness of the unified
metamodel abstractions, we were able to define changes that affect
aggregates, references and variations. The operations have been implemented
for three widely used NoSQL stores, one based in documents and schemaless,
other column-based that requires schema declarations and a third one based
in graphs. The usefulness of our proposal has been validated through a
refactoring of the StackOverflow dataset and an outlier migration on the
Reddit dataset. Also note that this work presents an application of the
unified metamodel presented in~\cite{metamodel2021}. An implementation of
Athena and Orion are publicly available on a GitHub
repository\footnote{\url{https://github.com/catedrasaes-umu/NoSQLDataEngineering}}.

Although the main purpose of the Orion language is to support schema
changes in a platform-independent way, it can be used in other
cases:~(i)~If no initial schema is provided, an Orion script can bootstrap
a schema by itself;~(ii)~Differences between Athena schemas may be
expressed as Orion specifications; and~(iii)~Orion specifications may be
obtained from specifications of existing tools such as the PRISM/PRISM++
operation language~\cite{curino-evolution2013}.

The future work considered includes:~(i)~Updating application code that
makes use of the retrieved data as well as handling query rewriting. Some
preliminary work has been done on~\cite{carlos-thesis2022}, where code
analysis is proposed to extract schemas, apply refactorings and provide
suggestions of code modifications, and this functionality could be
integrated in Orion.~(ii)~Investigating new operations to be added to the
taxonomy, such as operations regarding schema inheritance and type
hierarchies, and refining existing ones as needed.~(iii)~Extending Orion to
generate code for specific programming languages, which will allow to
implement operations on databases that are not supported natively.
Finally,~(iv)~integrating Orion into a tool for agile migration.

\bibliographystyle{plainurl}
\bibliography{ms}

\end{document}